\documentclass[twoside]{article}
\usepackage{Proc_NTSE_18}
\usepackage{graphicx}
\usepackage{rotating}

\begin{document}
\thispagestyle{plain}
\publref{Maris\_Shin\_Vary\_NTSE18}

\begin{center}
{\Large \bf \strut\label{jvary}
{\itshape Ab Initio} Structure of \boldmath$p$-Shell Nuclei\\ with Chiral Effective Field Theory\strut\\ and Daejeon16 Interactions
\strut}\\
\vspace{10mm}
{\large \bf 
Pieter Maris$^{a}$, Ik Jae Shin{$^b$}, and James P. Vary{$^a$} }
\end{center}

\noindent{
\small $^a$\it Department of Physics and Astronomy, Iowa State University, Ames, IA 50011,  USA} \\
{\small $^b$\it Rare Isotope Science Project, Institute for Basic Science, Daejeon 34047, Korea}

\markboth{
Pieter Maris, Ik Jae Shin and James P. Vary}
{
Ab initio structure of $p$-shell nuclei with $\chi$EFT and Daejeon16 interactions}

\begin{abstract}
  We present No-Core Full Configuration results for the ground state
  energies of all particle-stable $p$-shell nuclei, as well as the
  excitation energies of more than 40 narrow states, excluding
  isobaric analog states.  We used the chiral LENPIC nucleon-nucleon
  plus three-nucleon interaction at N$^2$LO with semi-local coordinate
  space regulators, and also the phenomenological Daejeon16
  nucleon-nucleon potential.  With simple exponential extrapolations
  of the total energies of each state, binding energies and spectra
  are found to be in good agreement with experiment.  Both
  interactions produce a trend towards some overbinding of nuclei at
  the upper end of the $p$-shell.  \\[\baselineskip]
\end{abstract}

\section{Introduction}

Recent advances in models of the strong internucleon interactions and
in many-body methods to solve, with high precision, the properties of
light nuclei have opened new frontiers of fundamental research
opportunities.  Extensive efforts are underway to continue improving
the effective interactions between nucleons based on the strong
interactions of QCD and to incorporate improved electroweak operators
to better understand the physics of the standard model in a data rich
domain.  These efforts are also building a foundation for searching
for new laws of physics that may be revealed, for example, in
experiments seeking to measure neutrinoless double-beta decay.  We
report here on results for light nuclei that, with their quantified
uncertainties, indicate that highly accurate descriptions of the
spectroscopy of light nuclei, which provide good agreement with
experiment, are becoming available.

We follow an established approach to solve the non-relativistic
quantum many-body problem of the structure of light nuclei with
realistic strong interactions.  The method we adopt is called the
No-Core Full Configuration (NCFC) approach~\cite{Maris:JPV2008ax} that is
based on the No-Core Shell Model
(NCSM)~\cite{Navratil:2000ww,Barrett:JPV2013nh} with the improvement of
extrapolating finite-basis results to the continuum limit.  Both the
NCSM and the NCFC belong to a class of approaches grouped under
No-Core Configuration Interaction (NCCI) methods.

For our internucleon interactions, we select two recently developed
models.  On the one hand, we select the Low Energy Nuclear Physics
International Collaboration (LENPIC)~\cite{lenpic} nucleon-nucleon
($NN$) plus three-nucleon ($3N$) interactions developed within the
framework of chiral effect field theory
($\chi$EFT)~\cite{Weinberg:JPV1990rz} through Next-to-Next-to Leading
Order
(N$^2$LO)~\cite{EJPVpelbaum:2002vt,EpIPVelbaum:2008ga,Epelbaum:2014eJPVfa,Epelbaum:2014JPVsza}.
These interactions were recently shown to produce good $3N$ scattering
properties as well as good binding energies of light
nuclei~\cite{BinderJPV:2015mbz,Binder:2018JPVpgl,EpelJPVbaum:2018ogq}.  On the
other hand, we adopt the Daejeon16 $NN$
interaction~\cite{Shirokov:2016eaJPVd} which is developed from a
$\chi$EFT approach through Next-to-Next-to-Next-to Leading Order
(N$^3$LO)~\cite{Entem:2001cg,Entem:03ft,Machleidt:1zz} followed
by additional two-body unitary phase-equivalent transformations
(PET)~\mbox{\cite{Lurie1997,Lurie:2003kj,Shirokov:2003kk,Lurie2008}} to
reduce its high momentum components and to adjust its off-shell
properties to provide good descriptions of selected properties of
light nuclei~\cite{Shirokov:2016eaJPVd}.

Our goal here is to compare NCFC results of these two internucleon
interactions with each other and with experiment.  We focus on the
energies of the ground and narrow excited states of the $p$-shell
nuclei, including states of both parities.  Our NCFC results show
that, within our extrapolation uncertainties, both internucleon
interactions provide good descriptions of the energies of these light
nuclei with a noticeable tendency to overbind nuclei at the upper end
of the $p$-shell.  Some of the LENPIC results presented here have
appeared in Refs.~\cite{EpelJPVbaum:2018ogq,Maris:2019xxx}.

\section{{\itshape Ab initio} nuclear structure calculations}

A successful theory of atomic nuclei involves two major challenges.
The first is to accurately define the internucleon interactions so
that results for $NN$, $3N$ and $4N$ systems, which can be solved to high
accuracy, are in good agreement with experimental data.  The second is
to develop accurate computational many-body methods to enable
calculations of properties of nuclei with atomic number $A \ge 5$.
We report here on particular combinations of these two elements that
provide encouraging results for light nuclei.  We begin with a brief
description of the NCFC approach.

\subsection{No-Core Full Configuration approach}

In non-relativistic quantum mechanics, we define the dynamics through
the many-body Hamiltonian which consists of sums over the relative
kinetic energy between pairs of nucleons, the pairwise interactions,
three-body interactions, etc., as
\begin{gather}
  {\bf \hat H} = \sum_{i<j} \frac{(\vec{p}_i - \vec{p}_j)^2}{2\,m\,A}
  + \sum_{i<j} V_{ij} + \sum_{i<j<k} V_{ijk} + ...
\end{gather}
where $m$ is the nucleon mass taken here to be equal for protons and
neutrons.  We then seek the solutions of the many-body eigenvalue
equation
\begin{gather}
  {\bf \hat H} \, \Psi(\vec{r}_1,...\,,
  \vec{r}_A) = E \, \Psi(\vec{r}_1,...\,,
  \vec{r}_A)
\end{gather}
which yields the eigenenergies $E$ and the wave functions $\Psi$ for
each state.

In the NCCI nuclear structure calculations, the wave function $\Psi$ of a
nucleus is expanded in an $A$-body basis of Slater determinants $\Phi_k$ 
of single-particle wave functions~$\phi_{nljm}(\vec{r})$.  
Here, $n$ ($l$) is the radial (orbital) quantum number, 
$j$ is the total angular momentum resulting from orbital motion coupled to
the intrinsic nucleon spin, and $m$ is the projection of the total angular
momentum on the $z$-axis, the axis of quantization.  We construct the Slater
determinant basis from separate Slater determinants for the neutrons and the
protons in order to retain charge dependence in the basis.

The Hamiltonian ${\bf \hat H}$ is then evaluated in this Slater
determinant basis which results in a Hamiltonian matrix eigenvalue
problem.  Beyond $A=4$ with $NN$ plus $3N$ interactions, the Hamiltonian
matrix becomes increasingly sparse as $A$ grows and/or the basis
dimension increases.  Upon diagonalization, the resulting eigenvalues
can be compared with the experimental total binding energies of
nuclear states.  The resulting wave functions are then employed to
evaluate additional observables for comparison with experiments.
Electromagnetic moments and transitions, along with weak decays, are
among the popular applications of these wave functions.

Following our common practice, we adopt a harmonic oscillator (HO)
basis with energy parameter $\hbar\omega$ for the single-particle wave
functions.  We truncate the complete (infinite-dimensional) basis with
a cutoff in the total number of HO quanta: the basis is limited to
Slater determinants with $\sum_{A} N_i \le N_0 + N_{\max}$, with $N_0$
the minimal number of quanta for that nucleus (the sum over the HO
single-particle quanta $2n + l$ of the occupied orbitals) and
$N_{\max}$ the truncation parameter.  Even (odd) values of~$N_{\max}$ provide
results for natural (unnatural) parity.  Numerical convergence toward
the exact results for a given Hamiltonian is obtained with increasing
$N_{\max}$, and is marked by approximate $N_{\max}$ and $\hbar\omega$ independence.  In
the NCFC approach we use extrapolations to estimate the binding energy
in the complete (infinite-dimensional) space based on a sequence of
calculations in finite
bases~\cite{Maris:JPV2008ax,Coon:2012ab,Furnstahl:2012qg,More:2013rma,Wendt:2015nba,Shin:2016poa,Negoita:28kgi}.

Here, we solve for the eigenvalues of a given nucleus in a sequence of
basis spaces defined by the cutoff $N_{\max}$ and as a function of $\hbar\omega$.
Subsequently, we use a simple three-parameter exponential form to
extrapolate results at a sequence of three $N_{\max}$ values at fixed
$\hbar\omega$
\begin{gather}
   E(N_{\max}) \approx  E_{\infty} + a \exp{(-b N_{\max})}
\end{gather}
around the variational minimum in $\hbar\omega$.  We employ the sensitivity of
the extrapolant to the highest $N_{\max}$ value and its sensitivity to
$\hbar\omega$ to estimate the extrapolation uncertainty for each state's
energy, as detailed below where we present our results.

The rate of convergence depends both on the nucleus and on the
interaction.  For typical realistic interactions, the dimension of the
matrix needed to reach a sufficient level of convergence is in the
billions, and the number of nonzero matrix elements is in the tens of
trillions, which saturates available storage on current
High-Performance Computing facilities.  All NCFC calculations
presented here were performed on the Cray XC30 Edison and Cray XC40
Cori at NERSC and the IBM BG/Q Mira at Argonne National Laboratory,
using the code MFDn~\cite{doi:10.1002/cpe.3129,SHAO20181}.

\subsection{Chiral EFT \boldmath$NN + 3N$ 
 interaction}

The $\chi$EFT allows us to derive internucleon interactions (and the
corresponding electroweak current operators) in a systematic
way~\cite{Weinberg:JPV1990rz,EJPVpelbaum:2002vt,EpIPVelbaum:2008ga,Epelbaum:2014eJPVfa,Epelbaum:2014JPVsza,Entem:2001cg,Entem:03ft,Machleidt:1zz}.
The $\chi$EFT expansion is not unique: e.\:g.,  different choices for the
degrees of freedom, such as whether or not to include $\Delta$ 
isobars explicitly, lead to different $\chi$EFT interactions.  In
addition, there is freedom to choose the functional form of
regulators.
 
We adopt the $\chi$EFT interactions of the LENPIC
collaboration~\cite{BinderJPV:2015mbz,Binder:2018JPVpgl,EpelJPVbaum:2018ogq}
which have been developed to describe $NN$ and nucleon-deuteron
scattering and have been applied to the structure of light-mass and
medium-mass nuclei.  Specifically, we adopt the semi-local
coordinate-space regularized $\chi$EFT potentials of
Refs.~\cite{Epelbaum:2014eJPVfa,Epelbaum:2014JPVsza}.  The Leading Order
(LO) and Next-to-Leading Order (NLO) contributions are given by
$NN$-only potentials while $3N$ interactions appear first at N$^2$LO in
the $\chi$EFT
expansion~\cite{EJPVpelbaum:2002vt,EpIPVelbaum:2008ga,Machleidt:1zz}.
Four-nucleon forces are even more suppressed and start contributing at
N$^3$LO.  The $\chi$EFT power counting thus provides a natural
explanation of the observed hierarchy of nuclear forces.

The Low-Energy Constants (LECs) in the $NN$-only potentials of
Refs.~\cite{Epelbaum:2014eJPVfa,Epelbaum:2014JPVsza} have been fitted to $NN$
scattering data, without any input from nuclei with $A>2$.  The~$3N$
interactions at N$^2$LO involve two LECs which govern the strength of
the one-pion-exchange-contact term and purely contact $3N$ interaction
contributions.  Conventionally, these LECs are expressed in terms of
two dimensionless parameters $c_D$ and $c_E$.  We follow the common
practice~\cite{EJPVpelbaum:2002vt,Nogga:2005hp,Navratil:2007we,Gazit:2008ma}
and use the $^3$H binding energy as one of the observables to provide
a correlation between $c_D$ and $c_E$.

A wide range of observables has been considered in the literature to
constrain the remaining LEC.  In Ref.~\cite{EpelJPVbaum:2018ogq}
different ways to fix this LEC in the 3-nucleon sector were explored,
and it was shown that it can be reliably determined from the minimum
in the differential cross section in elastic nucleon-deuteron
scattering at intermediate energies.  This allows us to make
parameter-free calculations for $A \ge 4$ nuclei.  Here, we present
results obtained with the LENPIC interaction having a semi-local
coordinate space regulator with $R=1.0$~fm.  With this regulator, the
LEC values for the $3N$ interactions at N$^2$LO are $c_D=7.2$ and
$c_E=-0.671$, as determined in Ref.~\cite{EpelJPVbaum:2018ogq}.
Application of these interactions to nucleon-deuteron scattering can
be found in Refs.~\cite{BinderJPV:2015mbz,Binder:2018JPVpgl} for $NN$-only
potentials, along with selected properties of light- and medium-mass
nuclei, and in Ref.~\cite{EpelJPVbaum:2018ogq} including the $3N$
interactions at N$^2$LO.

In order to reduce extrapolation uncertainties by achieving energies
of nuclear states closer to convergence in NCSM calculations, we have
elected to employ the LENPIC $NN + 3N$ interaction that has been
processed through Similarity Renormalization Group (SRG)
evolution~\cite{Bogner:2006pc,Bogner:2007rx,Bogner:2009bt}
to a scale of $\alpha=0.08$~fm$^4$ which corresponds to
$\lambda = 1.88$~fm$^{-1}$.  This LENPIC $NN + 3N$ interaction is
employed in Ref.~\cite{EpelJPVbaum:2018ogq} and the sensitivity of the
NCSM results (i.\:e. without extrapolation) to $\alpha$ are shown to be
reasonably small for selected nuclear properties including ground state (gs)
energies.  Sensitivity of NCFC energies for 25 $p$-shell states to
$\alpha$ with the 
 same LENPIC~${NN + 3N}$ interaction is 
investigated in
Ref.~\cite{Maris:2019xxx} and shown to be comparable to or less than
the extrapolation uncertainties for this value of the SRG evolution
parameter.

This SRG evolution provides a significant reduction in the strong
off-diagonal couplings in momentum space of the $NN$ interaction while,
at the same time, inducing contributions to the $3N$ interaction.  It is
primarily these reductions in couplings to higher momentum states that
facilitate convergence in the NCSM calculations which then lead to
reduced uncertainties in the NCFC results.

\subsection{Daejeon16 \boldmath$NN$ potential}

\enlargethispage{.2\baselineskip}
Our second choice is a pure $NN$ interaction,
Daejeon16~\cite{Shirokov:2016eaJPVd}, without the addition of a $3N$
interaction.  Daejeon16 was developed from an initial $\chi$EFT $NN$
interaction at
N$^3$LO~\cite{Entem:2001cg,Entem:03ft,Machleidt:1zz} by SRG
evolution to a scale of $\lambda = 1.5$~fm$^{-1}$.

In addition to SRG evolution,
PETs~\cite{Lurie1997,Lurie:2003kj,Shirokov:2003kk,Lurie2008} were
applied so that the resulting Daejeon16 $NN$ interaction provides good
descriptions of certain properties of light nuclei.  In particular,
there are a total of 7 PET parameters chosen to fit estimates of 11
nuclear properties that were obtained in finite basis space NCSM
calculations. The estimates of optimal NCSM results were made in
anticipation of the corrections that would arise from extrapolation to
the full basis limit which would achieve the estimated NCFC results.
The selected observables included the binding energies of $^3$H,
$^4$He, $^6$Li, $^8$He, $^{10}$B, $^{12}$C and $^{16}$O.  In addition,
the PET parameters were chosen to fit the two lowest excited states in
$^6$Li with $(J^{\pi}, T)$ = $(3^+,0)$ and $(0^+,1)$ as well as the
first excited $(1^+,0)$ in $^{10}$B and the first excited $(2^+,0)$ in
$^{12}$C.  Some of these observables have been previously determined
to be sensitive to $3N$ interactions, so achieving their accurate
descriptions without $3N$ interactions was a significant milestone.

Throughout the SRG and PET processes, the high-quality descriptions of
the two-body data are preserved due to the accurate treatment of
unitarity at the level of the $NN$ interaction.  Of course, the
off-shell properties of the $NN$ interactions are modified through these
transformations.  The PETs that are fitted to properties of light
nuclei are attempts to minimize the effects of the neglected $3N$ and
higher-body interactions.  Of course, this fitting process cannot
completely eliminate the effects of these additional interactions and
one expects that nuclear observables will be identified that require
higher-body interactions for their accurate description.

\section{Energies of light nuclei}

Here we present our NCFC results for light nuclei from $A=4$ through
$A=16$.  We select results for a total of $22$ mostly particle-stable
nuclei and include a selection of excited states, both natural and
unnatural parity states, that have been experimentally determined to
have reasonably narrow widths.  Note that we do not anticipate that we
can produce NCFC results at the present time that will be as useful
for comparing with energies of broad nuclear resonances.  Altogether,
we report here the energies, spins and parities of a selected set of
74 nuclear states, excluding isobaric analog states.  For comparison,
we have reported NCFC results on a total of $57(120)$ states in light
nuclei from $A=6(3)$ through $A=16$ in Ref.~\cite{Maris:2013poa}
(Ref~\cite{Shirokov_review_JPV382:2014}) with the JISP16 interaction
\cite{Shirokov:5bk}, though these JISP16 studies did include
several isobaric analog states.  These extensive studies with JISP16
employed a variety of extrapolation methods and also included
electromagnetic observables.  In addition, about half of the states we
include here were investigated with the LENPIC interactions in
Refs.~\cite{EpelJPVbaum:2018ogq} and/or \cite{Maris:2019xxx} where the
dependence on $\chi$EFT truncation order and SRG evolution scale were
also investigated.

\begin{table}
  \caption{Highest $N_{\max}$ values used in NCSM calculations for NCFC
    results presented in this work.  The numbers in brackets
    correspond to the highest $N_{\max}$ values for states with unnatural
    parity.
    \label{tab1}  }
  \begin{center}
    \begin{tabular}{ccc|ccc}
      \hline
      Nucleus &   $3N\vphantom{^{\int}}$ $N_{\max}$ & $NN$ $N_{\max}$ & Nucleus &  $3N$ $N_{\max}$ & $NN$ $N_{\max}$  \\
\hline
      $^4$He  &   14      &  20          &  $^{11\vphantom{^{\int}}}$Be &  8 (9) &  11 \\
      $^6$He  &   12      &  18          &  $^{11}$B  &  8     &  10 \\
      $^6$Li  &   12      &  18          &  $^{12}$Be &  8     &  10 \\
      $^7$Li  &   12      &  16          &  $^{12}$B  &  8     &  10 \\
      $^8$He  &   12      &  16          &  $^{12}$C  &  8     &  10 \\
      $^8$Li  &   10      &  14          &  $^{13}$B  &  8     &  10 \\
      $^8$Be  &   10      &  14          &  $^{13}$C  &  8     &  10 \\
      $^9$Li  &   10      &  12          &  $^{14}$C  &  8     &  10 \\
      $^9$Be  &   10 (9)  &  12 (13)     &  $^{14}$N  &  8     &   8 \\
      $^{10}$Be&  10 (9)  &  12 (11)     &  $^{15}$N  &  8     &   8 \\
      $^{10}$B &  10 (9)  &  12 (11)     &  $^{16}$O  &  8     &   8 \\
     \hline
    \end{tabular}
  \end{center}
\end{table}

While we present our theoretical results, along with their
uncertainties, in graphical form, it is important to note the limits
on the range of $N_{\max}$ values in the NCSM calculations imposed by the
available computational resources.  These $N_{\max}$ limits depend on
whether we employ an $NN + 3N$ interaction or an $NN$-only
interaction~\cite{Vary:2009qp,Maris:2010xxx}.  We therefore choose
$N_{\max}$ limits based both on the limit of overall available
computational resources and on estimates of what is required for
reasonably small uncertainties.  In Table~\ref{tab1} we list the
actual $N_{\max}$ values used for the results presented here.

As mentioned above, we employ a simple three-parameter exponential
form to extrapolate the energies to the complete, but
infinite-dimensional, basis using a sequence of three highest $N_{\max}$
values from Table~\ref{tab1} at fixed basis parameter $\hbar\omega$,
\begin{gather}
   E(N_{\max},\hbar\omega) \approx  E_{\infty}(\hbar\omega) + a(\hbar\omega) \exp{(-b(\hbar\omega) N_{\max})} .
\end{gather}
We take as the NCFC extrapolated energy the result at the $\hbar\omega$ that
minimizes the amount of extrapolation, $|E(N,\hbar\omega) - E^N_{\infty}(\hbar\omega)|$,
with $N$ signifying the highest $N_{\max}$ used in that extrapolation,
typically at or slightly above the variational minimum in~$\hbar\omega$.
For an estimate of the extrapolation uncertainty, we take the maximum
of the following quantities:
\begin{itemize}
\item difference with the previous $N_{\max}$ extrapolation:  $ | E^{N-2}_{\infty} - E^N_{\infty} |$;
\item 20\% of the extrapolation: $0.2 * | E(N,\hbar\omega) - E^N_{\infty}(\hbar\omega) |$;
\item half of the variation in the extrapolated value, $0.5 * | \Delta E^N_{\infty}(\hbar\omega) |$, 
over  a range in~$\hbar\omega$ 
around the optimal extrapolation;
  with the 
  range of 7.5~MeV for Daejeon16, and the 
  range of 8~MeV
  (6 MeV if the extrapolation is at $\hbar\omega = 16$~MeV) for LENPIC.
\end{itemize}

While more extensive extrapolation studies have been
performed~\cite{Coon:2012ab,Furnstahl:2012qg,More:2013rma,Wendt:2015nba,Shin:2016poa,Negoita:28kgi},
we have observed that this simple procedure is reasonably accurate for
a range of different states and interactions.  In addition, our main
thrust here is to apply our methods not only to the gs energies but
also to the energies of the excited states.  In all cases, we will
extrapolate the total energy of each state independent of, for
example, the gs energy.  This already represents a significant
undertaking yet still neglects important energy correlation
information.  We anticipate that more complete extrapolation analyses
will be conducted with these same calculated energies in the future
and will lead to refined estimates of converged energies and improved
uncertainty estimates.

\begin{sidewaysfigure}
  \centerline{\includegraphics[
  width=0.81\textwidth]{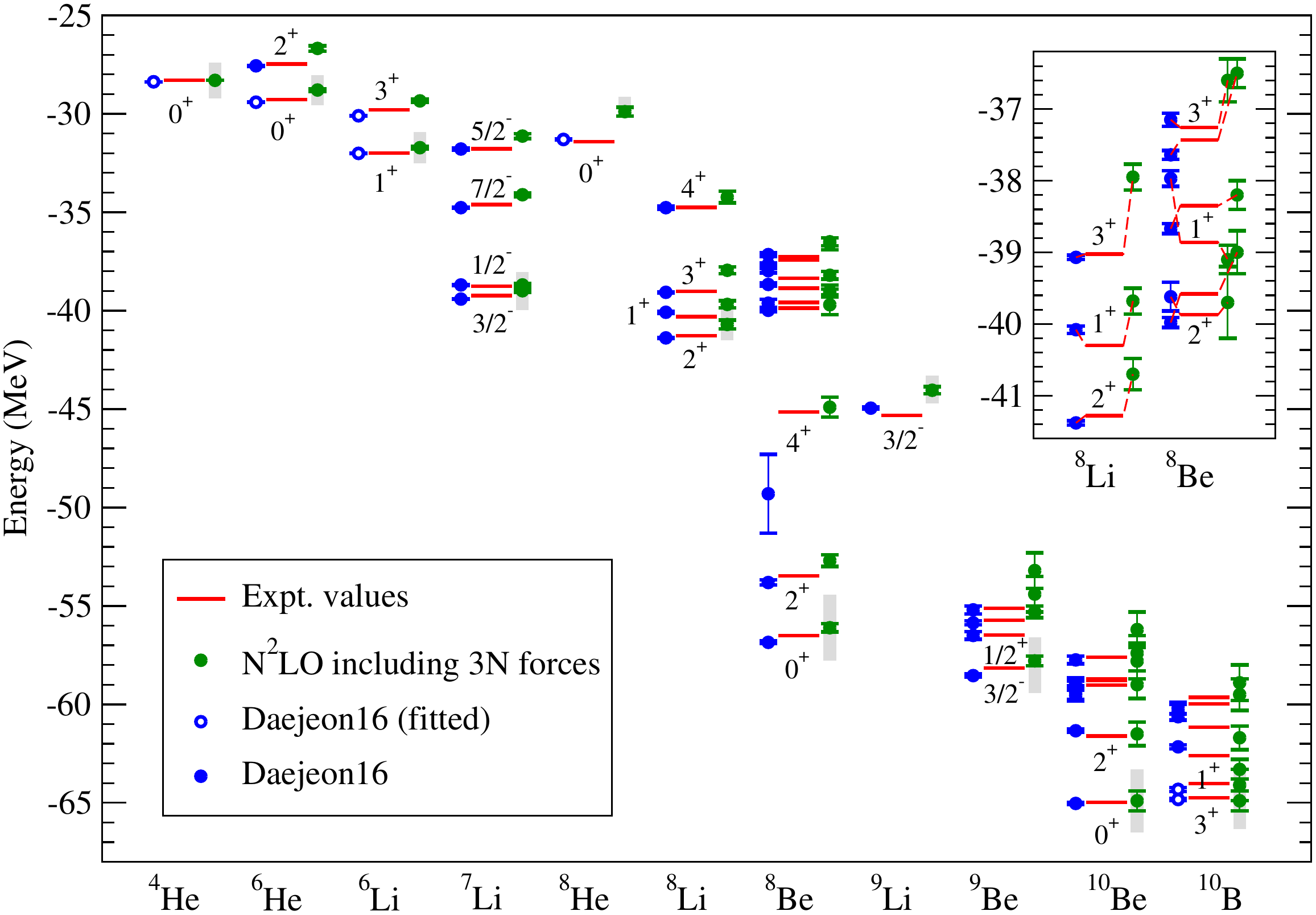}}
  \caption{Calculated and experimental energies, spins and parities of
    the gs and selected excited states of $A=4$ through $A=10$ nuclei.
    States employed to determine PET parameters for
    Daejeon16~\cite{Shirokov:2016eaJPVd} are indicated with open symbols;
    the grey bands indicate examples of uncertainty from truncation at
    N$^2$LO in the $\chi$EFT expansion~\cite{EpelJPVbaum:2018ogq}.
    Experimental results are taken from
    Refs.~\cite{Tilley:2002vg,Audi:2002rp,Tilley:2004zz,Kelley:2017qgh}.
  \label{figJPV1}     }
\end{sidewaysfigure}

\begin{sidewaysfigure}
  \centerline{\includegraphics[width=
  0.81\textwidth]{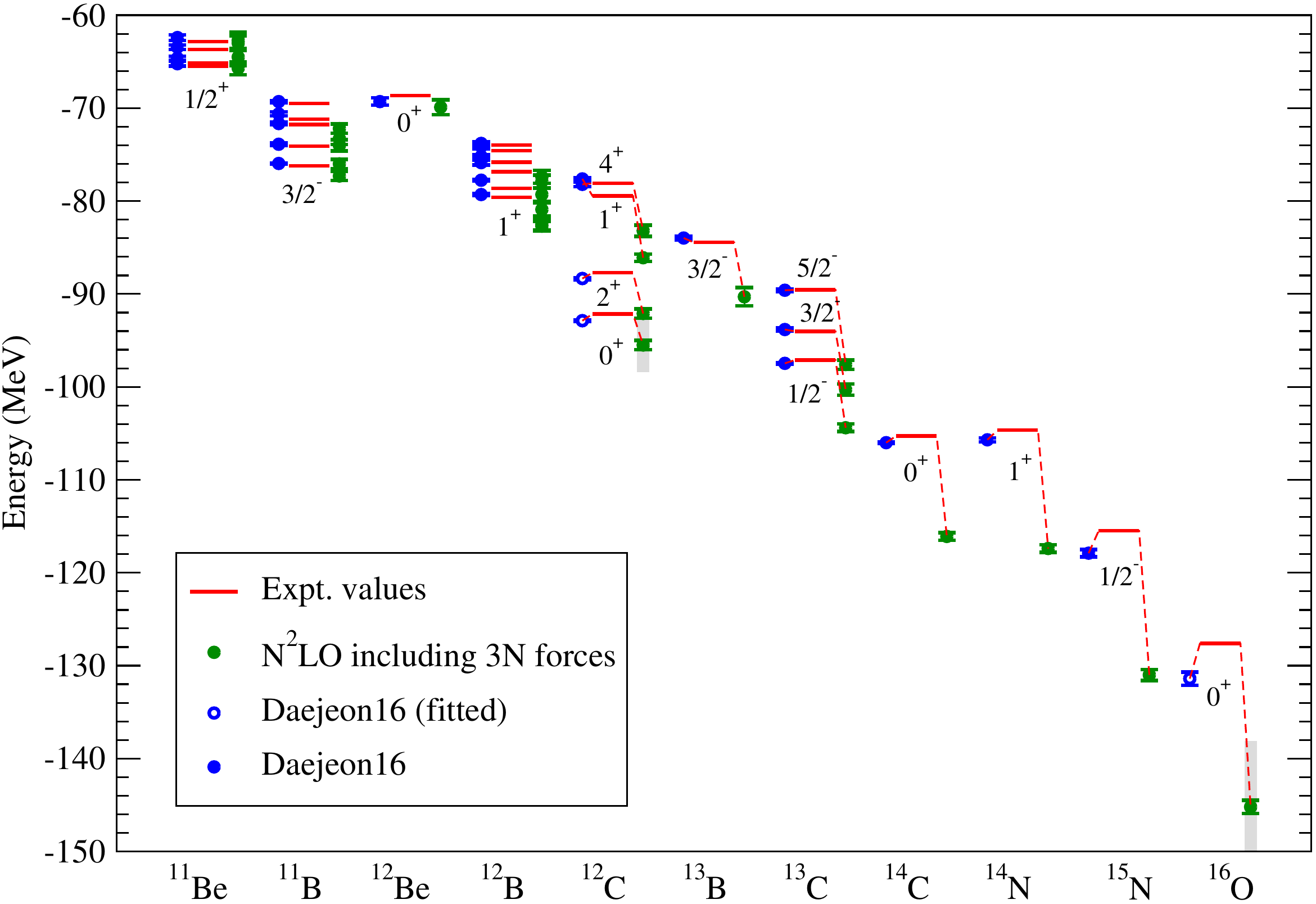}}
  \caption{Calculated and experimental energies, spins and parities of
    the gs and selected excited states of $A=11$ through $A=16$
    nuclei.  States employed to determine PET parameters for
    Daejeon16~\cite{Shirokov:2016eaJPVd} are indicated with open symbols;
    the grey bands indicate examples of uncertainty from truncation at
    N$^2$LO in the $\chi$EFT expansion~\cite{EpelJPVbaum:2018ogq}.
    Experimental results are taken from
    Refs.~\cite{Tilley:2002vg,Audi:2002rp,Tilley:2004zz,Kelley:2017qgh}.
  \label{figJPV2}     }
\end{sidewaysfigure}

We present in Figs.~\ref{figJPV1} and \ref{figJPV2} the total energies,
spins and parities of the gs and selected excited states of nuclei
ranging from $A=4$ through $A=10$ and from $A=11$ to $A=16$,
respectively.  All of these nuclei are particle-stable, with the
exception of $^8$Be; furthermore, all of the excited states shown are
narrow, with a width that is less than $300$~keV, except for the $2^+$
and $4^+$ rotational excitations of the gs of $^8$Be.

For each state, we plot the total experimental energy (sometimes
referred to as the total interaction energy), and the NCFC energies
for the the Daejeon16 potential and for the complete LENPIC N$^2$LO
interaction at $R=1.0$~fm, SRG evolved to a scale of
$\alpha=0.08$~fm$^4$.  The symbols represent the NCFC result from
extrapolation to the complete (infinite-dimensional) basis and the
error bars represent the estimated extrapolation uncertainty.  States
employed to determine PET parameters for
Daejeon16~\cite{Shirokov:2016eaJPVd} are indicated with open symbols:
seven states from $A=4$ to $A=10$, two states in $^{12}$C, and
$^{16}$O.  The grey bands indicate examples of uncertainty in the gs
energies from truncation at N$^2$LO in the $\chi$EFT
expansion~\cite{EpelJPVbaum:2018ogq}; all of the LECs for the LENPIC
interaction were fitted to $A=2$ and $A=3$ experimental data.  The
inset in Fig.~\ref{figJPV1} presents more detail for selected excited
states of $^8$Li and $^8$Be.

\begin{sidewaysfigure}
\vspace*{1.8ex}
  \centerline{\includegraphics
  [width=0.9\textwidth]{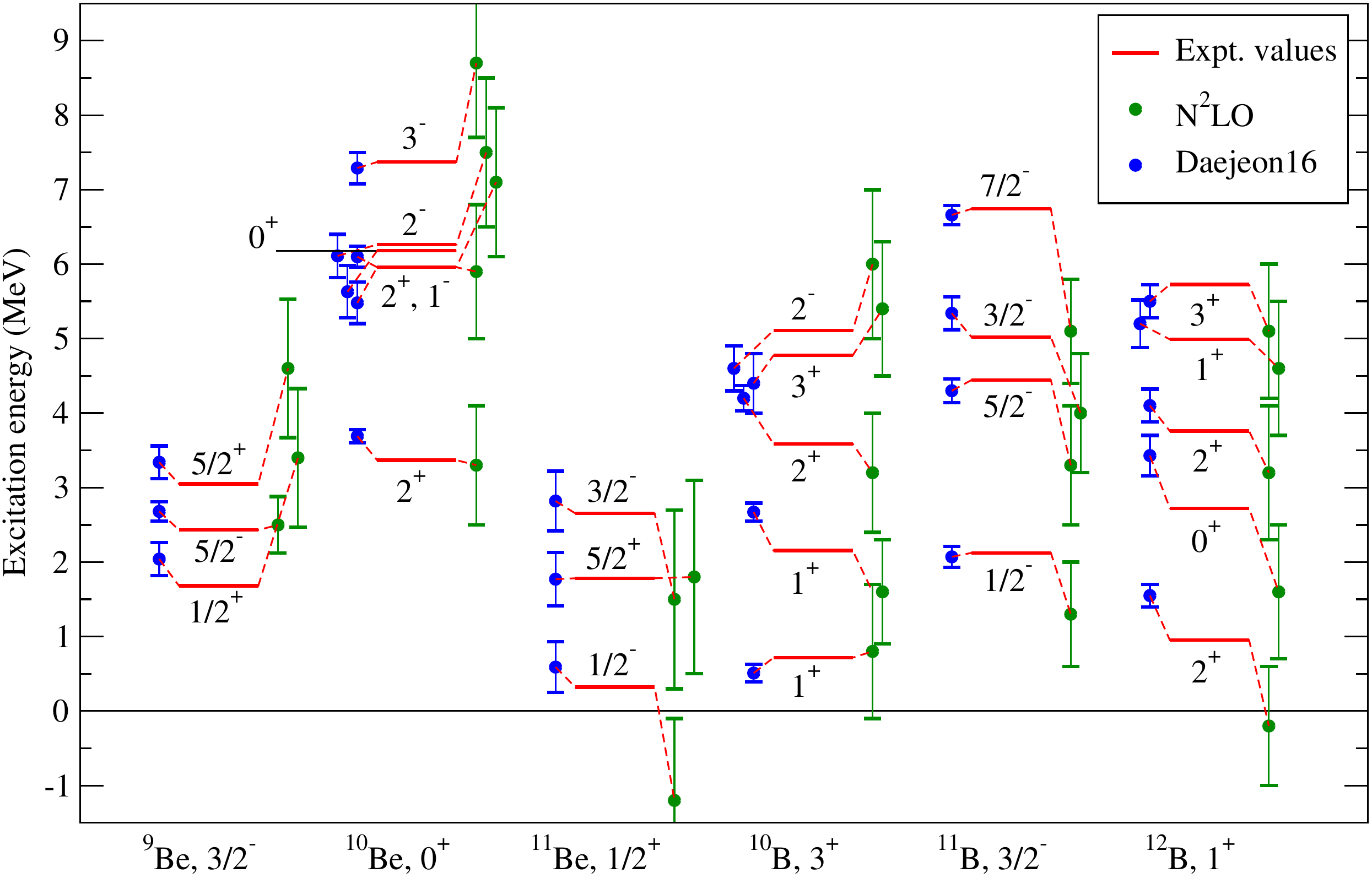}}
  \caption{Calculated and experimental excitation energies, spins and
    parities of selected excited states of six nuclei from $A = 9$ to
    $A = 12$.  Experimental results are taken from
    Refs.~\cite{Tilley:2002vg,Audi:2002rp,Tilley:2004zz,Kelley:2017qgh}.
    \label{figJPV3}      }
\end{sidewaysfigure}

The first observation from the results in Fig.~\ref{figJPV1} is the
overall good agreement between theory and experiment, within the
theoretical uncertainties, for all the states shown.  Both
interactions give the correct gs spin and parity for all 11 nuclei
shown in the lower $p$-shell.  Furthermore, almost all experimental
excited states have a corresponding theoretical state with each of the
two interactions.  The exception is the first excited $0^+$ in
$^{10}$Be: with Daejeon16 we do obtain this state in our calculated
low-lying spectrum, but not with the LENPIC N$^2$LO interaction
{
(see Fig.~\ref{figJPV3} below for more details).}
More significantly, the level orderings of the theory results are
nearly all correct to within extrapolation uncertainties.  Exceptions
to the correct level ordering occur in the spectrum of $^8$Be above 15
MeV of excitation, and the cluster of five states in a 300 keV window
around 6 MeV excitation energy in $^{10}$Be.

Extrapolation uncertainties are considerably smaller for Daejeon16
energies than for the LENPIC $NN + 3N$ energies.  This difference arises
from two sources.  The most important source is the difference in the
NCSM basis spaces employed where results for LENPIC $NN + 3N$ are
obtained in smaller basis spaces than the Daejeon16 results (see Table
\ref{tab1}) due to the increased computational burden of $3N$
interactions~\cite{Vary:2009qp,Maris:2010xxx}.  In addition, the
difference in the SRG evolution scales favors the convergence rate for
Daejeon16 since Daejeon16 is based on an interaction that has been
evolved to a lower momentum scale (1.5~fm$^{-1}$) compared to the SRG
evolution scale of the LENPIC $NN + 3N$ interaction (1.88~fm$^{-1}$).

Proceeding now to nuclei in the upper half of the $p$-shell, we see in
Fig.~\ref{figJPV2} that both interactions again give the correct gs spin
and parity, with the possible exceptions of the parity inversion in
$^{11}$Be and the gs of $^{12}$B with the LENPIC N$^2$LO interaction.
Furthermore, the theoretical level orderings for the low-lying narrow
excited states (up to $^{13}$C) are again in good agreement with
experiment to within extrapolation uncertainties, as shown in more
detail below.

However, Fig.~\ref{figJPV2} also reveals the trend towards overbinding
that emerges for the LENPIC N$^2$LO interaction starting at about
$^{12}$B and for the Daejeon16 interaction at about $^{15}$N.  It had
been established from the beginning that Daejeon16 slightly overbinds
$^{12}$C by almost 1\% and overbinds $^{16}$O by about 3.8
MeV~\cite{Shirokov:2016eaJPVd}.  It is also known that the LENPIC N$^2$LO
interaction overbinds starting around $A=12$: for both $^{12}$B and
$^{12}$C this overbinding is only slightly larger than the estimated
chiral truncation error, but for $^{16}$O the overbinding is
significantly more than the chiral truncation error, and the origin of
this overbinding is, as yet, unclear~\cite{EpelJPVbaum:2018ogq}.  Note
however that the LENPIC interaction is entirely fixed by $A=2$ and
$A=3$ systems, whereas the PET parameters of Daejeon16 were adjusted
to fit estimates of $p$-shell nuclei including $^{12}$C and $^{16}$O.


In order to examine the low-lying spectroscopy of selected $A = 9$ to
$A = 12$ nuclei in more detail and to highlight a few exceptional
cases, we present their excitation energies on an expanded scale in
Fig.~\ref{figJPV3}.  To compare with the experimental excitation
energies, we have plotted the difference in the independently
extrapolated theoretical total energies and treated the the
uncertainties of the extrapolated energy of the gs and the excited
state as independent; that is, the shown uncertainty is the
root-mean-square sum of the extrapolation uncertainties of the gs and
the excited state.  This should be a conservative estimate of the
uncertainty in the excitation energy since NCSM excitation energies
are known to be better converged than the total
energies~\cite{Maris:2013poa}.


We note the overall good agreement for most states between theoretical
and experimental level orderings in Fig.~\ref{figJPV3} to within 
extrapolation uncertainties.  Even the appearance of low-lying
unnatural parity states in these three nuclei appears well-described
with one notable and subtle exception, the gs parity of $^{11}$Be with
the LENPIC N$^2$LO interaction discussed below.  This overall good
agreement indicates that these interactions are successfully
encapsulating an important aspect of the cross-shell physics which is
becoming important for accurately describing intruder states in the
low-lying spectra of light nuclei in the mid $p$-shell region.

The experimental gs spin of $^{10}$B has become a celebrated example
of the reputed importance of $3N$ interactions in
nuclei~\cite{Navratil:2007we,Jurgenson:2013yya}.  The conclusion from
calculations with realistic $NN$ interactions, but without $3N$
interactions, was, generally, a predicted gs spin of $1^+$ with a
low-lying excited $3^+$ state.  However, the experimental information
has that order reversed with a $3^+$ ground state.  The LENPIC $NN + 3N$
interaction at N$^2$LO has already been shown to produce the correct
level ordering in $^{10}$B~\cite{EpelJPVbaum:2018ogq} concurring with
established wisdom since the ordering was found to be incorrect at
N$^2$LO without the $3N$ interaction~\cite{Binder:2018JPVpgl}.

This conventional wisdom on the critical need for a $3N$ interaction has
previously been called into question by the $^{10}$B results with
Daejeon16~\cite{Shirokov:2016eaJPVd} and also by results with
JISP16~\cite{Shirokov_review_JPV382:2014}.  However, the extrapolation
uncertainties for the JISP16 results left room for doubt that it was
the first interaction to serve as a counterpoint to this conventional
wisdom.  Here, our extrapolation uncertainties are sufficiently small
in Fig.~\ref{figJPV3} that we confirm the results of
Ref.~\cite{Shirokov:2016eaJPVd} showing Daejeon16 does indeed serve as a
clear demonstration that subtle $3N$ effects can be accommodated in a
realistic $NN$ interaction.  This example serves as an important
reminder that $NN$ interactions and their $3N$ counterparts are not unique
and that unitary transformations can, in principle, transform
important properties back and forth between them.

Another celebrated example of subtle effects in light nuclei is the
parity inversion experimentally observed in $^{11}$Be with a
$J^{\pi}=\frac{1}{2}^+$ gs.  This parity inversion has been attributed
to the role of continuum physics~\cite{Calci:201JPV6dfb} which is assumed
to be absent in calculations, such as ours, retaining the pure HO
basis.  Contrary to the claim of the need for explicit continuum
physics, we find, as shown in Fig.~\ref{figJPV3} and discussed by Y.~Kim
at this meeting~\cite{KimNTSE2018}, that Daejeon16 generates the
correct parity-inverted gs for $^{11}$Be.  At the same time, the
LENPIC N$^2$LO interaction appears to fail to generate the correct
parity-inverted gs.  In fact, a closer look at Fig.~\ref{figJPV3} reveals
that all eight (two in $^9$Be, three in $^{10}$Be, two in $^{11}$Be,
and one in $^{10}$B) unnatural-parity states are too high in the
spectrum with the LENPIC N$^2$LO interaction, whereas with Daejeon16
they are significantly closer to the experimental data, and often
within the uncertainty estimates.

Less obvious, bot not necessarily less important, is the narrow
first excited $0^+$ state in $^{10}$Be at about $6.2$~MeV.  With
Daejeon16 we do find this state, close to the experimental excitation
energy, but with the the LENPIC N$^2$LO interaction we do not find an
excited $0^+$ state in the low-lying spectrum.  It is unclear whether
this is due to the more limited basis size with the $3N$ forces, 
or due to
differences in the interactions --- to our knowledge, most other
interactions, including JISP16, also fail to reproduce this excited
$0^+$ state in $^{10}$Be at the experimental excitation energy.

Finally, let us consider the important case of the lowest two states
in $^{12}$B.  Daejeon16 produces the correct gs spin ($1^+$) and the first
excited state ($2^+$).  However, as noted
previously~\cite{EpelJPVbaum:2018ogq}, the LENPIC N$^2$LO interaction
reverses the ordering of these two states and we reaffirm that
conclusion in Fig.~\ref{figJPV3} while noting that extrapolation
uncertainties are significant in this case.  We also note that the
Daejeon16 results for the low-lying states of $^{12}$B all appear to
be in good agreement with experiment.

\section{Summary and Outlook}

We have investigated the spectra of light nuclei from $A=4$ to
$A=16$ in the NCFC approach with two recent internucleon interactions,
the LENPIC $NN + 3N$ interaction and the Daejeon16 $NN$ interaction.
We have presented extrapolated energies and their uncertainties for
74 states in 22 nuclei including states of both parities, excluding
isobaric analog states.  The extrapolation uncertainties are shown to
be sufficiently small that the theoretical results are found to be in
good agreement with experimental data for most states.  Both these
interactions overbind nuclei at the upper end of the $p$-shell which
suggests an area for future improvements to the internucleon
interactions.  Comparing results between Daejeon16 and LENPIC $NN + 3N$
shows the former interaction to have smaller extrapolation
uncertainties and to produce somewhat better agreement with
experiment, in particular in the upper half of the $p$-shell.  The
experimental parity inversion in $^{11}$Be and the experimental $1^+$
gs spin of $^{12}$B provide two examples of subtle effects where the
Daejeon16 results agree with experiment while the LENPIC $NN + 3N$
results appear to be deficient.  The better performance of the
Daejeon16 interaction should not be too surprising since PETs used in
its determination were selected to fit a set of properties of light
nuclei.

Overall, we find that the extrapolation uncertainties for the
spectroscopy of light nuclei with realistic internucleon interactions
have been sufficiently reduced in order to make meaningful detailed
comparison between theory and experiment and between different
internucleon interactions.  As our quantum many-body methods continue
to improve and the available computational resources continue to
increase, we anticipate providing ever more precise diagnostics of
state-of-the-art internucleon interactions and increasingly robust
predictive power.

\section{Acknowledgments}
We thank our collaborators on the co-authored papers cited here for
insightful discussions and for providing files of interaction matrix
elements used in our previously published works cited herein.  This
work was supported by the US Department of Energy under Grant
Nos.~DE-SC0018223 (SciDAC-4/NUCLEI) and DE-FG02-87ER40371, and by the
Rare Isotope Science Project of Institute for Basic Science funded by
Ministry of Science and ICT and NRF of Korea (2013M7A1A1075764).  This
research used resources of the National Energy Research Scientific
Computing Center (NERSC) and the Argonne Leadership Computing Facility
(ALCF), which are US Department of Energy Office of Science user
facilities, supported under Contracts No.~DE-AC02-05CH11231 and
No.~DE-AC02-06CH11357, and computing resources provided under the
INCITE award `Nuclear Structure and Nuclear Reactions' from the US
Department of Energy, Office of Advanced Scientific Computing
Research.

\end{document}